\documentclass[10pt,showpacs,twocolumn]{revtex4}
\usepackage{graphicx}
\usepackage{amsmath}
\usepackage{amssymb}
\usepackage{amsmath}
\usepackage{}

\begin{document}
\title{Tomographic reconstruction of molecular orbitals with twofold mirror antisymmetry: overcoming the nodal plane problem}
\author{Xiaosong Zhu,$^{1}$ Meiyan Qin,$^{1}$ Yang Li,$^{1}$ Qingbin Zhang,$^{1,2}$ Zhizhan Xu,$^{1,2}$ and Peixiang Lu$^{1,2}$ \footnote{Corresponding author: lupeixiang@mail.hust.edu.cn}}

\affiliation{$^1$ Wuhan National Laboratory for Optoelectronics and School of Physics, Huazhong University of
Science and Technology, Wuhan 430074, China\\
$^2$ Key Laboratory of Fundamental Physical Quantities Measurement of MOE, Wuhan 430074, China}
\date{\today}

\begin{abstract}
We propose a new method to overcome the nodal plane problem for the tomographic reconstruction of molecular orbitals with twofold mirror antisymmetry in the length form based on high-order harmonic generation. It is shown that, by carrying out the reconstruction procedure in the rotating laboratory frame using the component of the dipole moment parallel to the electron recollision direction, the nodal plane problem is avoided and the target orbital is successfully reconstructed. Moreover, it is found that, the proposed method can completely avoid the additional artificial lobes found in the results from the traditional method in the velocity form and therefore provides a more reliable reproduction of the target orbital.
\end{abstract} \pacs{32.80.Rm, 42.65.Ky} \maketitle

\noindent The fast progress in strong-field physics has made it possible to observe the structure and ultrafast electron dynamics in atoms and molecules with {\AA}ngstr\"{o}m and attosecond resolutions \cite{Lein,Lin,Smirnova,Pfeiffer,Wu,Zhou}. In the recent decade, a fascinating application known as the molecular orbital tomography (MOT) based on high-order harmonic generation (HHG) is raised \cite{Itatani,Zwan,Haessler,Hijano,Vozzi,Zwan2,Qin}. With the MOT scheme, a two-dimensional projection of the molecular orbital on the plane orthogonal to the pulse propagation direction can be imaged after measuring the HHG spectra from different alignments of the molecule. In detail, by calibrating the measured HHG signal of the target molecule with that of a reference system with the same ionization energy, the recombination dipole matrix element of the molecular orbital is extracted. Then based on the Fourier slice theorem, the target molecular orbital can be reconstructed by performing the inverse Fourier transform of the recombination dipole matrix elements (see Refs. \cite{Haessler2,Salieres,Diveki} for more information).

The reconstruction can be in principle performed in both the length form (LF) and the velocity form (VF). However, the applicabilities of the two forms are different. It is generally suggested to apply the velocity form if the target orbital has one or more nodal planes containing the coordinate axes in the molecular frame, because one would run into numerical problems in the length form \cite{Haessler2,Salieres,Diveki}. In this work, we show that this problem can be avoided if the orbital is antisymmetric with one nodal plane, and the problem only holds for the orbitals with twofold mirror antisymmetry. Accordingly, we propose a new method to overcome the nodal plane problem in reconstructing this kind of orbitals in the length form. Taking a $\pi_g$ orbital as an example, we compare the reconstructed result with that obtained from the traditional method in the velocity form, which shows that the proposed method provides a more reliable reproduction of the target orbital. Finally, we discuss the scope of application of this method and show that it can be extended to more types of twofold antisymmetric orbitals to overcome the nodal plane problem.

The scheme of MOT is illustrated in Fig. 1. (xyz) is the molecular frame and (x'y'z') is the laboratory frame. The probe laser pulse propagates along the $z'$ axis and is linearly polarized parallel to the $x'$ axis, along which the continuum electron recollides with momentum $k$ to generate the HHG radiation. $\theta$ is the alignment angle.

\begin{figure}[htb]
\centerline{\includegraphics[width=5.5cm]{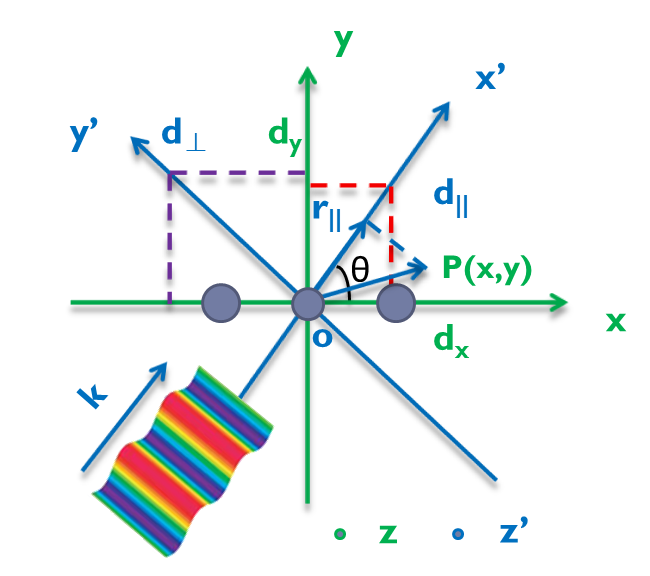}}
\setlength{\abovecaptionskip}{-0.1cm}
\setlength{\belowcaptionskip}{-0.15cm}
\caption{(Color online) Illustration of the MOT scheme: (xyz) is the molecular frame and (x'y'z') is the laboratory frame. The probe laser pulse propagates along the $z'$ axis and is linearly polarized parallel to the $x'$ axis, along which the continuum electron recollides with momentum $k$ to generate the HHG radiation. $\theta$ is the alignment angle of the molecule.}
\end{figure}

In the length form, the traditional reconstruction method can be expressed as \cite{Haessler2,Salieres,Diveki}:
\begin{eqnarray}
&\Psi_x^{LF}(x,y)=\frac{{\mathcal{F}}_{k\rightarrow r}[d_x(k_x,k_y)]}{x},\\
&\Psi_y^{LF}(x,y)=\frac{{\mathcal{F}}_{k\rightarrow r}[d_y(k_x,k_y)]}{y},
\end{eqnarray}
where $\mathcal{F}_{k\rightarrow r}$ denotes the inverse Fourier transform. Since the reconstruction is derived in the molecular frame, the $x$ and $y$ components of the dipole moment ($d_x$ and $d_y$) are required. In experiments, the HHG radiation in both the $x'$ and $y'$ directions are measured first to extract $d_\|$ and $d_\bot$ using a polarizer formed by a pair of silver mirrors \cite{Haessler2,Levesque}, then $d_x$ and $d_y$ are calculated by projecting $d_\|$ and $d_\bot$ onto the $x$ and $y$ axes respectively: $d_x=d_\|\cos(\theta)-d_\bot\sin(\theta)$, $d_y=d_\|\sin(\theta)+d_\bot\cos(\theta)$. In the Fourier space, $k_x=k\cos(\theta)$, $k_y=k\sin(\theta)$ and $k^2/2=\omega$ according to the energy conservation, where $\omega$ is the harmonic frequency. Although the same orbital can in principle be reconstructed from both the $x$ and $y$ components of the dipole moments, they will most likely not give the same result due to the limited discrete sampling in Fourier space. Therefore, the final reconstructed orbital is generally defined as $\Psi_{xy}^{LF}(x,y)=\frac{1}{2}[\Psi_x^{LF}(x,y)+\Psi_y^{LF}(x,y)]$ to hopefully average out the distortions.

In Eqs. (1) and (2), if the target orbital has a nodal plane containing the $y$ or $x$ axis, one will run into numerical problems when dividing by $x$ or $y$ respectively \cite{Haessler2,Salieres,Diveki}. In detail, if the orbital has a nodal plane containing the $y$ axis for example (i.e. is antisymmetric with respect to the $y$ axis), the result of the inverse Fourier transform in the sampled spectral range in Eq. (1) $\mathcal{F}_{k\rightarrow r}[d_x]$ equals a nonzero value at x=0 and one will confront the numerical problem when performing the division by x since $\lim\limits_{x\rightarrow0}\frac{{\mathcal{F}}_{k\rightarrow r}[d_x]}{x}=\infty$. On the other hand, if the orbital is symmetric with respect to the $y$ axis (i.e. without the nodal plane), $\mathcal{F}_{k\rightarrow r}[d_x]\rightarrow0$ and the division $\frac{{\mathcal{F}}_{k\rightarrow r}[d_x]}{x}$ tends to a finite value for $x\rightarrow0$, so one will not run into the nodal plane problem. These results are due to the symmetry property of the orbitals and the feature of the inverse Fourier transform in the limited k-space. Besides the above discussed orbitals with symmetry or antisymmetry, the orbital can also be nonsymmetrically distributed, the reconstruction of which was discussed in \cite{Zhu}.

In previous works, it is generally suggested that one should employ the velocity form to avoid the nodal plane problem. Actually, this problem can also be avoided by choosing appropriate component of the dipole moment in the length form, if the orbital has only one nodal plane. Take the highest occupied molecular orbital (HOMO) of acetylene for example \cite{Vozzi2}, which is a $\pi_u$ orbital with only one nodal plane containing the $x$ axis as shown in Fig. 2(a). The target orbital can be reconstructed by using only $d_x$ (according to Eq. (1)) as shown in Fig. 2(b), while can not be reconstructed using $d_y$ (according to Eq. (2)) as shown in Fig. 2(c). The unexpected structure observed along the $x$ axis in Fig. 2(c) is just owing to the nodal plane problem. In Fig. 2(d), we also present the reconstructed result by using the traditional method in the velocity form \cite{Zwan,Haessler2}. Comparing panel (b) with (d), it is found that the target orbital is better reconstructed in the length form. In the simulation, the dipole moment is calculated by $\mathbf{d}=\langle\Psi|\mathbf{r}|e^{i\mathbf{k}\cdot\mathbf{r}}\rangle$, where $\Psi$ is the Hartree-Fock (HF) orbital obtained from the Gaussian 03 code \cite{Gaussian}. Throughout this work, the spectrum is sampled by the odd harmonics of the 800-nm laser field in the range of harmonics 17--51 \cite{Itatani}. The alignment angle $\theta$ is scanned from $0^\circ$ to $350^\circ$ in step of $10^\circ$. We assume a perfect alignment of the molecules in the simulation and the molecules are considered to be fixed in space during the interaction.

\begin{figure}[htb]
\setlength{\abovecaptionskip}{-0.1cm}
\setlength{\belowcaptionskip}{-0.15cm}
\centering\includegraphics[width=6.8cm]{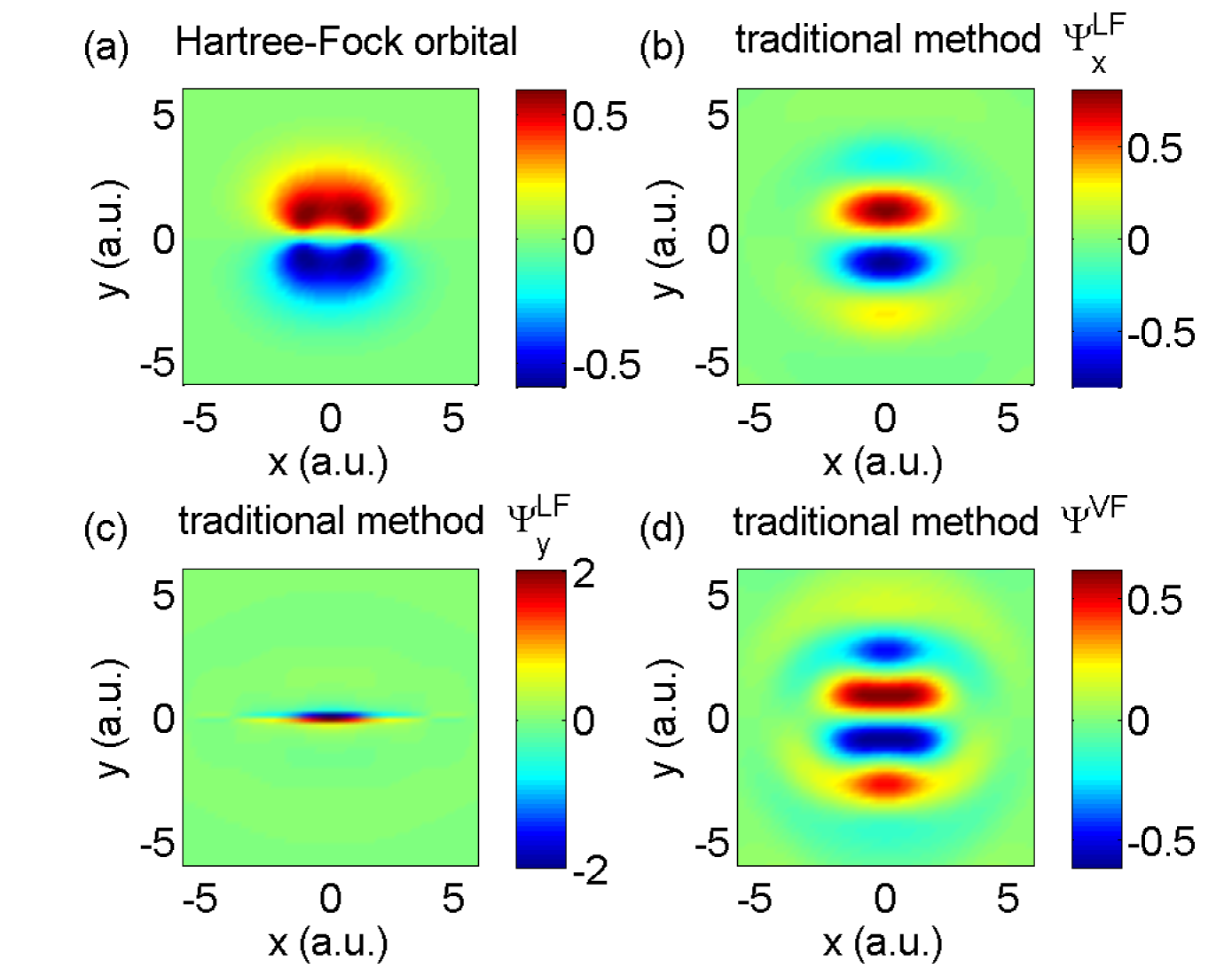} \caption{(Color online) 2D projections of the HOMO of acetylene obtained (a) from HF calculation, (b) according to Eq. (1), (c) according to Eq. (2), and (d) with the traditional method in VF.}
\end{figure}

However, when the target orbital has two nodal planes containing the $x$ and $y$ axes respectively (\emph{i.e.} with the twofold mirror antisymmetry), one will run into the numerical problem in both Eqs. (1)(2) and the nodal plane problem remains unavoidable. Take the HOMO of CO$_2$ with $\pi_g$ symmetry for example \cite{Vozzi}. Fig. 3(a) shows the HF orbital obtained from the Gaussian 03 code, in which two nodal planes divide the orbital into four lobes with alternating signs. Fig. 3(b) shows the reconstructed result with the traditional method in the length form. What one can see is only the cross-shaped structure resulting from the nodal plane problem.

\begin{figure}[htb]
\centering\includegraphics[width=6.8cm]{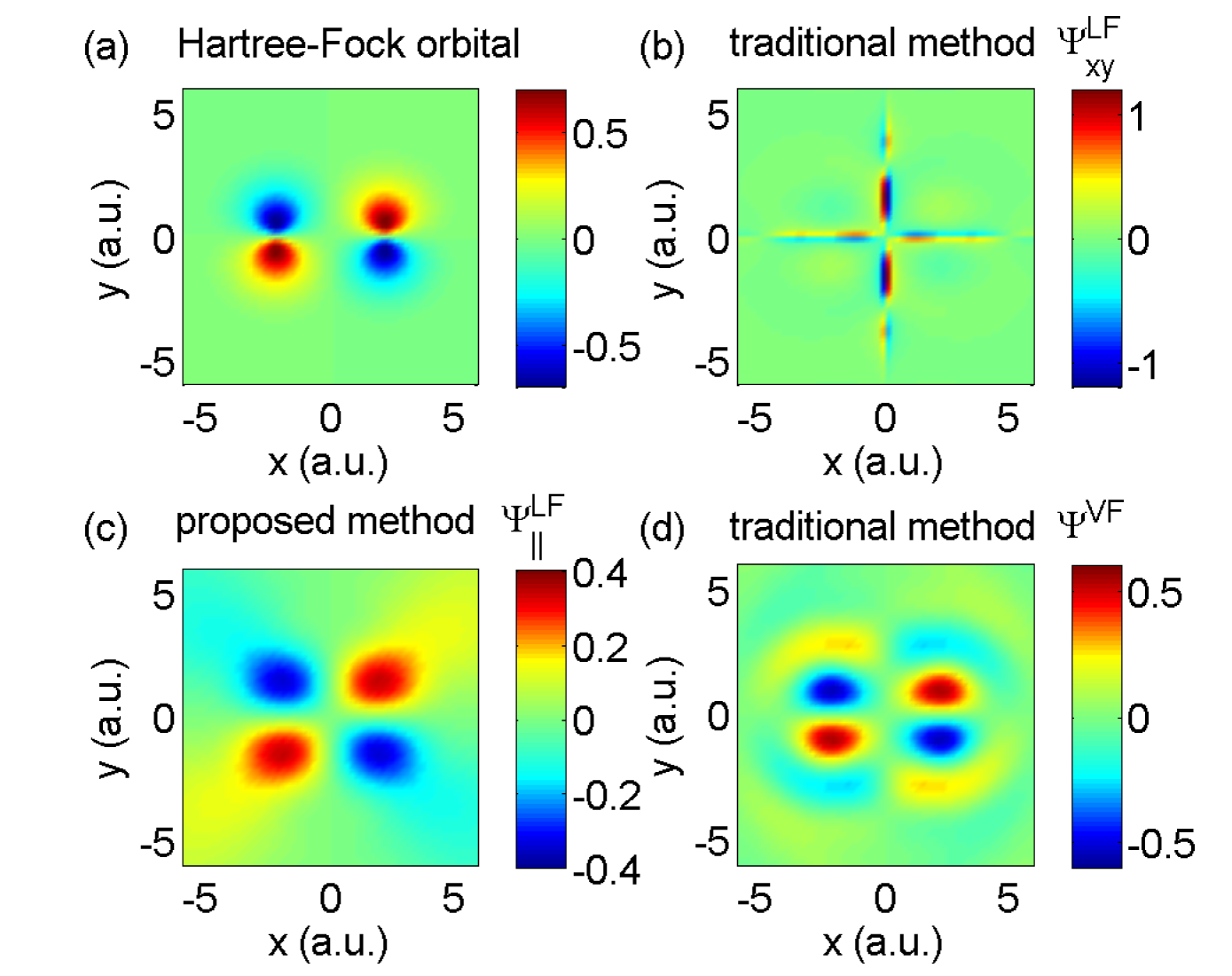}
\setlength{\abovecaptionskip}{-0.1cm}
\setlength{\belowcaptionskip}{-0.15cm}
\caption{(Color online) 2D projections of the HOMO of CO$_2$ obtained (a) from HF calculation, (b) with the traditional method in LF, (c) with the proposed method, and (d) with the traditional method in VF.}
\end{figure}

To overcome this problem, we propose a new reconstructing method in the length form. The basic idea is to reconstruct the orbital using $d_{\parallel}$ directly in the laboratory frame instead of projecting $d_{\parallel}$ and $d_{\bot}$ at different $\theta$ onto the fixed $x$ and $y$ axes in the molecular frame. According to $d_{\parallel}(k,\theta)=\langle\Psi|r_{\parallel,\theta}|e^{i\mathbf{k}_{\theta}\cdot\mathbf{r}}\rangle$, the target orbital is reconstructed following
\begin{equation}
\Psi_{\parallel}^{LF}(x,y)=\sum_{\theta}\frac{{\mathcal{F}}_{k\rightarrow r}[d_{\parallel,\theta}(k_x,k_y)]}{r_{\parallel,\theta}(x,y)}.\\
\end{equation}
For any point P(x,y), $r_{\parallel,\theta}(x,y)=x\cos(\theta)+y\sin(\theta)$ is the projection of the vector $\overrightarrow{OP}$ onto the $x'$ axis (as shown in Fig. 1). In this method, the division is performed in the rotating direction $\mathbf{r}_{\parallel,\theta}$ (signal along $\mathbf{r}_{\parallel,\theta}$ divided by the coordinate on $\mathbf{r}_{\parallel,\theta}$) rather than in the fixed $x$ and $y$ directions perpendicular to the nodal planes. In this frame, the inverse Fourier transform can be expressed as
\begin{align}
\mathcal{F}_{k\rightarrow r}[d_{\parallel,\theta}]&=\int d_{\parallel,\theta}\exp[ik(x\cos\theta+y\sin\theta)]dk\\
=\int &d_{\parallel,\theta}\cos(kr_{\parallel,\theta})dk+i\int d_{\parallel,\theta}\sin(kr_{\parallel,\theta})dk
\end{align}

To explain why the proposed method can avoid the nodal plane problem in a concise form, we assume that the molecular orbital is real-valued. (An arbitrary complex phase of the orbital does not change the result of the reconstruction and can actually be set to zero by applying a global phase shift \cite{Zwan,Salieres}.) In this case, if the target orbital $\Psi$ is symmetric about the origin of the molecular frame, $d_{\parallel,\theta}$ is purely imaginary-valued \cite{Haessler}. Therefore, the reconstructed orbital should be
\begin{align}
Re[\Psi_{\parallel}^{LF}]&=\sum_{\theta}\frac{Re[{\mathcal{F}}_{k\rightarrow r}(d_{\parallel,\theta})]}{r_{\parallel,\theta}}\\
\propto &\sum_{\theta}\sum_k\frac{id_{\parallel,\theta,k}\sin(kr_{\parallel,\theta})}{r_{\parallel,\theta}}.
\end{align}
According to Eq. (7), since the limitation of the division $\lim\limits_{r_{\parallel,\theta}\rightarrow0}\frac{id_{\parallel,\theta,k}\sin(kr_{\parallel,\theta})}{r_{\parallel,\theta}}=ikd_{\parallel,\theta,k}$ equals a finite value for each alignment angle $\theta$, the nodal plane problem is avoided.

The reconstructed result with this proposed method for CO$_2$ is shown in Fig. 3(c). The main structure of the HOMO of CO$_2$ is successfully reproduced, and the main error is only the diffusion of the orbital in the outer region. As a comparison, we also present the result reconstructed by the traditional method in the velocity form in Fig. 3(d). In this figure, one could see eight or more lobes in the reconstructed result (the same as in \cite{Haessler2,Vozzi}). The additional artificial lobes would be misleading for people to learn the structure of an ``unknown'' orbital. This problem is also found in Fig. 2(d): the reconstructed orbital from the traditional method in the velocity form is seriously distorted by the additional lobes.

To compare the results more clearly, the slices of the HF orbital, the reconstructed orbitals $\Psi_{\parallel}^{LF}$ and $\Psi^{VF}$ of the HOMO of CO$_2$ are presented in Fig. 4. We compare two groups of slices. Panels (a) and (b) show the cuts along the two diagonal directions respectively, and panels (c) and (d) depict the cuts along $y=1\ a.u.$ and $x=2\ a.u.$ respectively. The comparison shows that, although the slices of $\Psi^{VF}$ match better with those of the HF orbital for the amplitudes of the peaks, they have serious oscillations which do not exist in the slices of the HF orbital. These oscillations just correspond to the additional lobes. On the other hand, regarding $\Psi_{\parallel}^{LF}$ reconstructed by the proposed method, although the result does not match the HF orbital exactly for the absolute values and converge to zero more slowly, it completely avoids the oscillations (\emph{i.e.} the additional lobes). The comparison indicates that, without the misleading additional lobes, the proposed method provides a more reliable reproduction of the target orbital. It is also worth noting that, the qualities of the reconstructed results are also affected by the choice of the sampled spectral range for both the traditional method in velocity form and the proposed method. Since the width of the detected spectrum is limited in experiments, the sampled spectral range should be carefully chosen. This has been systematically discussed in \cite{Haessler2}, and it is summarized that the spectral range should contain a characteristic spatial frequency of the orbital and a ``well balanced'' amount of the positive and negative spectra amplitudes. In this work, the reconstructed results are expected to be better for both reconstructing methods, if the spectral range is further extended. However, one might confront more experimental problems to achieve the broader detected spectral range.

\begin{figure}[htb]
\centering\includegraphics[width=6.5cm]{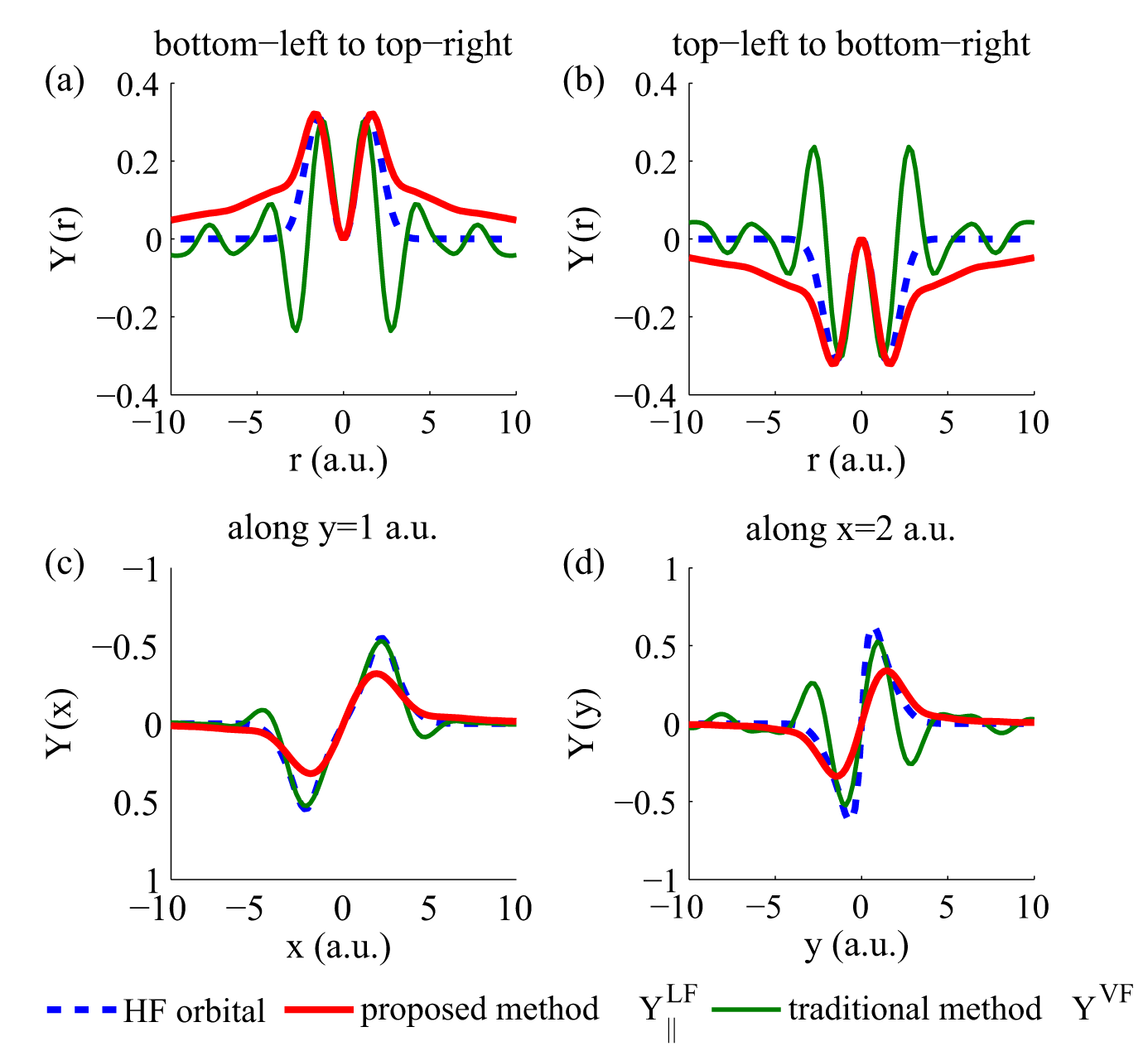}
\setlength{\abovecaptionskip}{-0.1cm}
\setlength{\belowcaptionskip}{-0.15cm}
\caption{(Color online) Comparison of the slices of the HF orbital, the reconstructed orbitals $\Psi_{\parallel}^{LF}$ and $\Psi^{VF}$ for CO$_2$: (a)(b) along the diagonal directions, and (c)(d) along two straight lines parallel to the $x$ and $y$ axes respectively.}
\end{figure}

Before conclusion, we would like to briefly discuss the scope of application of the proposed method. Although the applicability of this method is not limited by the nodal planes of the target orbital, it can not be used for all kinds of orbitals either. It requires the target orbital to be symmetric about the origin of the molecular frame, or one will again run into numerical problems when dividing by $r_{\parallel}$ according to Eqs. (5--7) and related discussion. For the orbitals with twofold antisymmetry, which have the nodal plane problem, this requirement is satisfied. Therefore, this method can be applied to various types of molecular orbitals with the twofold antisymmetry to overcome the nodal plane problem.

In summary, the nodal plane problem can be avoided for the antisymmetric orbitals with one nodal plane, and only holds for the orbitals with twofold antisymmetry for the traditional method in the length form. To overcome this problem, we propose a new method by performing the reconstruction procedure in the rotating laboratory frame using the component $d_{\parallel}$ parallel to the electron recollision direction. With this method, the HOMO of CO$_2$ is successfully reconstructed and we further compare the reconstructed result with that from the traditional method in the velocity form. The comparison shows that, our proposed method completely avoids the additional artificial lobes found in the orbitals obtained in the velocity form and therefore provides a more reliable reproduction of the target orbital. Finally, it is discussed that this method can be extended to more types of orbitals with the twofold antisymmetry.

This work was supported by the NNSF of China under grants 11234004 and 60925021, the 973 Program of China under grant 2011CB808103, and the Doctoral fund of Ministry of Education of China under grant 20100142110047.

\end{document}